\documentclass[dvips]{article}

\usepackage{icrc2011}

\title{Probing the CR positron/electron ratio at few hundreds GeV through Moon shadow observation with the MAGIC telescopes}

\newcommand{\etal}{\MakeLowercase{\textit{et al. }}} 
\shorttitle{Colin \etal Moon shadow observation with MAGIC}

\authors{Pierre Colin$^{1}$,
 Daniela Borla Tridon$^{1}$,
 Alicia Diago Ortega$^{2}$,
 Marlene Doert$^{3}$,
 Michele Doro$^{4}$,
 Jonathan Pochon$^{2}$,
 Nikola Strah$^{3}$,
 Tihomir Suri\'c$^{5}$,
 Masahiro Teshima$^{1}$
 on behalf of the MAGIC collaboration
}
\afiliations{$^1$Max-Planck-Institut f\"ur Physik, Munich, Germany\\
 $^2$Instituto de Astrofisica de Canarias, La Laguna, Spain\\
 $^3$Universit\"at Dortmund, Germany\\
 $^4$Universitat Aut\'onoma de Barcelona, Spain\\
 $^5$Institute R. Boskovic, Croatia }
\email{Mailto: colin@mppmu.mpg.de}

\abstract{ The antimatter components measured in the Cosmic Ray (CR) flux are thought as
secondary particles induced by the propagation of galactic CRs within the galaxy.
Recent results from the PAMELA experiment show an unexpected increase of the positron
electron ratio above 10 GeV. There could be different interpretations to explain that
result, the most discussed ones being the signature of nearby compact astrophysical
source(s) or of dark matter annihilation/decay. Probing the positron-fraction rise
above 100 GeV would help to disentangle among different scenarios. Imaging Atmospheric
Cherenkov Telescopes (IACT) can extract the cosmic lepton signal from the hadronic CR
background between a few hundred GeV and a few TeV and reconstruct energy and
incident direction with a very good resolution. In addition, by using the natural
spectrometer formed by the Moon and the geomagnetic field, it is possible to measure
the positron/electron ratio at the TeV regime through the observation of the CR Moon
shadow. Despite the technique is particularly challenging because of the high background
light induced by the Moon and the treatment of data, the MAGIC collaboration has performed
for the first time such observations in 2010 and 2011. Here we present the observation strategy
and the performance achieved during this campaign.}

\begin{document}
\maketitle

\section{Introduction}
The propagation of cosmic rays (CRs) in the galaxy induces secondary
particles by interaction with the interstellar medium. The relative
abundance of different CR components ($\bar p$/$p$, B/C), as well as
the diffuse $\gamma$-ray emission at the GeV regime, strongly
constrain the CR propagation models\,\cite{Strong04}. Generally, these
models predict a smooth all-electron (e$^+$+\,e$^-$) spectrum decreasing
faster than a power law $E^{-3}$ above a few tens of GeV and a
positron fraction e$^+$/(e$^+$+\,e$^-$) decreasing slowly with the
energy. However, recent measurements show a different picture (see
Fig~1). The all-electron spectrum measured with
\textit{Fermi}-LAT\,\cite{Abdo09}, ATIC\,\cite{Chang08},
H.E.S.S.\,\cite{Aharonian08} and MAGIC\,\cite{Borla11} is harder than
expected above 30\,GeV with a break around 800\,GeV. In the ATIC data
the feature is even more pronounced with a significant bump around
500\,GeV. Furthermore, PAMELA\,\cite{Adriani09, Mocchiutti11}
reported an increasing positron fraction above
10\,GeV, in agreement with the previous results of HEAT\,\cite{Beatty04} and AMS-01\,\cite{Aguilar07},
and recently confirmed by \textit{Fermi}-LAT\,\cite{Vandenbroucke11}. 

 \begin{figure}[!t]
  \centering
  \includegraphics[width=6.5cm]{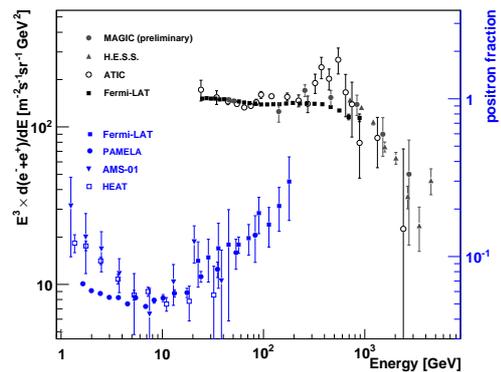}
  \caption{CR (e$^+$\,+\,e$^-$) spectrum \cite{Abdo09, Chang08,
      Aharonian08, Borla11} 
   and the positron fraction (e$^+$/(e$^+$\,+\,e$^-$))
   \cite{Adriani09, Beatty04, Aguilar07, Vandenbroucke11}.}
  \label{fig1}
 \end{figure}

The anomalies in the e$^+$ and \,e$^-$ fluxes are generally interpreted
as the presence of a new component with a harder spectrum and a higher
$e^+$/$e^-$ ratio than the fluxes expected by classical CR models.
Because of the short lifetime of TeV electrons in the galaxy, this extra-component
must come from nearby sources ($<1-2$\,kpc). Many scenarios involving dark matter
(annihilation/decay), pulsars or modified CR propagation models have
been proposed to interpret the data\,\cite{Fan10}. The positron
fraction predicted by these models above 100\,GeV can be very
different. Measurement of this ratio at higher energies is thus
essential to discriminate between models and to establish a connection
between the $e^+$ fraction rise and the all-electron bump.

In the near future, PAMELA, AMS-02\,\cite{AMS02} or \textit{Fermi}-LAT may
extend at higher energy their e$^+$ fraction measurement by
collecting more data but more likely they should not reach much higher
energies than 300\,GeV. Using the Moon shadow effect, Imaging
Atmospheric Cherenkov Telescope (IACT) experiments could measure or
constrain the $e^+$/$e^-$ ratio around 1\,TeV. The idea of probing the
$e^+$ fraction by the observation of the Moon shadow has been
proposed at the last ICRC\,\cite{Colin09}. Since then, the MAGIC
collaboration has developed this new detection technique and collected
several hours of data in stereoscopic mode. In the next section, the
principle of the Moon shadow effect is explained. Then, in section 3,
the Moon shadow observation strategy with MAGIC and the data analysis
method are presented. Finally, the performance of such method 
estimated from MC simulation on Crab Nebula data taken in similar
experimental conditions (high Moon light level), is discussed in
section~5, followed by concluding remarks. 

\section{The Earth/Moon spectrometer}
The Earth/Moon system forms a natural spectrometer in which the Moon
absorbs a part of the CRs creating a ``hole'' in the isotropic flux
(so called the Moon shadow) and the Earth magnetosphere deflects the
trajectory of any coming particle depending on its charge and momentum
(equivalent to its energy for an ultra-relativist particle). Then, the
position of the Moon shadow in the sky (observed from ground) is
different for each CR. For neutral CRs (like diffuse $\gamma$\,rays),
it lies at the actual Moon position. For charged CRs, the Moon shadow
is shifted perpendicularly to the geomagnetic field along an axis
close to an East-West orientation. Negative and positive CRs are
shifted respectively eastward and westward. The amplitude of 
the deviation depends on the particle rigidity. For ultra relativistic
particles, it is simply proportional to its charge $Z$ and inversely
proportional to its energy $E$. The typical deviation for medium
zenith angle observation ($\sim$45$^\circ$) is about $1.5^\circ \times Z
\times1\,TeV/E$.

The Moon shadow effect is used by ground-based EAS detectors and
cosmic neutrino experiments to estimate their angular resolution and
pointing accuracy \cite{DiSciascio11, Boersma11, Riviere11}. As most
of the CRs are positive particles (atom nuclei), the all-CR Moon
shadow is asymmetric with a larger deficit at the west side of the
Moon. EAS experiments with the lowest energy threshold have detected
this East-West asymmetry and derived upper limits on the $\bar{p}/p$
ratio ($\sim5-10$\%) at the TeV regime \cite{l3c2004, tibet2007,
  DiSciascio11}. However the poor electron/proton discrimination of
these experiments do not allow them to extract the electron signal
from the hadron CR background.

In contrast, IACT experiments are able to distinguish electromagnetic
showers (induced by $\gamma$ rays and electrons) and hadronic showers
with a very good accuracy. H.E.S.S.\cite{Aharonian08} and MAGIC\,\cite{Borla11}
measured the all-electron spectrum from 100\,GeV to a few TeV.
Therefore, in principle, they could also probe the $e^+$/$e^-$ ratio in this energy range
using the Moon shadow effect. However, the Moon shadow lies only a few degrees from the Moon
and then the dazzling background light induced by the scattered
moonlight must be handled properly. The first try to observe the
Moon-shadow effect with IACT was performed in the 90's with the
ARTEMIS experiment\,\cite{artemis2001}, which was using UV--filter
($200-300$\,nm) to suppress most of the moonlight. Unfortunately,
the Cherenkov light of the air showers was strongly suppressed too,
and the Moon shadow could not be detected.

Here, with MAGIC, we are using a different approach which do not include
major hardware modifications. All the coming photons are recorded and then
the background light is excluded during the image cleaning stage. 

\section{Observation with MAGIC}
MAGIC is a pair of 17\,m IACT located at the Canary Island of La 
Palma, 2200\,m above sea level, built for very high energy $\gamma$-ray astronomy.
The stereoscopic system has been in operation since fall 2009.
It generally operate during dark nights when it reaches its lower energy threshold
($\sim$50\,GeV) and a sensitivity above $\sim$300\,GeV of 0.8\% of the Crab Nebula in 50\,h
\cite{Carmona11}. The MAGIC camera uses low gain PMTs with only 6 dynodes, covering 3.5$^\circ$ field of view, that can
be operated at high background light levels without any damage. MAGIC
observes astrophysical objects under moderate moonlight for long time.
The sensitivity above 200\,GeV is almost unaffected by an increase of the Night Sky Background (NSB)
5 times higher than a dark-night sky\,\cite{Britzger09}. 

The Moon shadow observation with MAGIC is performed on the energy
range $300-700$\,GeV, where an excess in the all-electron spectrum is
measured, and where the e$^+$ fraction is unknown (see figure~1). The
position of the electron Moon shadow at such energies ranges lies
between 2$^\circ$ and 6$^\circ$ from the Moon. Thus we point the
MAGIC telescopes at about 4$^\circ$ from the Moon. As the Moon shadow
spreads out in only one direction (East-West) it can be contained in
one half of the camera. The other half is then used for the background
estimation. Figure~2 shows the typical observation of the Moon shadow with the
MAGIC telescopes. The camera center tracks a point 0.6$^\circ$ away
from the targeted electron shadow position ($400-500$\,GeV) in a
direction perpendicular to the Moon shadow deviation axis. The
telescopes point alternatively (every 10\,min) at each side of the
Moon shadow axis for a better control of the systematics (wobble observation).

 \begin{figure}[!t]
  \centering
  \includegraphics[width=6.5cm]{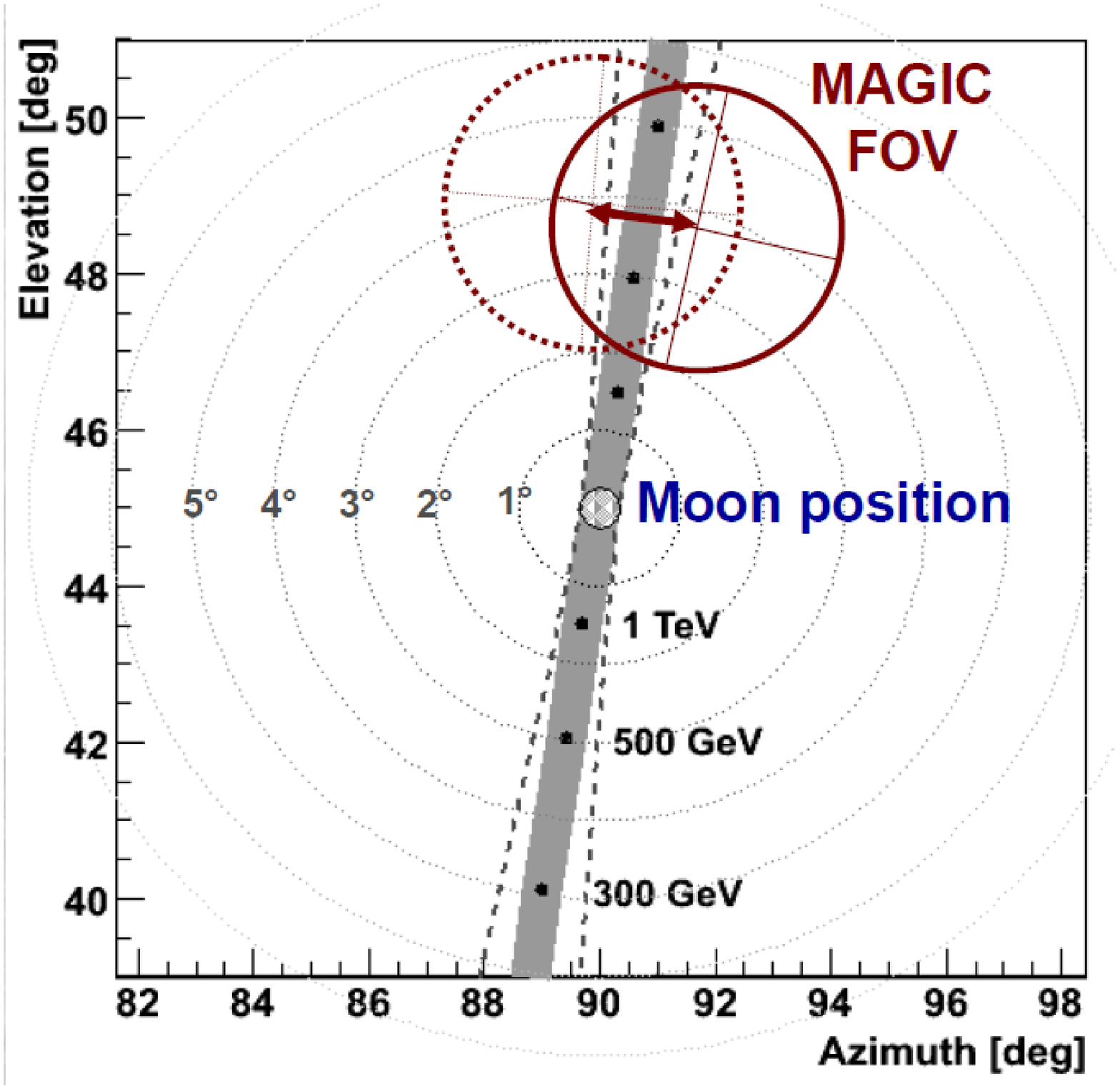}
  \caption{Positions of the Moon shadows for an observer at the MAGIC
    site with a rising Moon at 45$^\circ$ of elevation.  The $e^-$
    shadow is below the Moon (Eastward) and the $e^+$ shadow above the
    Moon (Westward). The dashed lines represent the position
    uncertainty induced by a 10$\%$ error on the geomagnetic
    field. The light-gray dotted circles are the curves of
    iso-distance to the Moon. The red circles show the MAGIC field of
    view during $e^+$ shadow observation (wobble mode) of the energy
    range $300-700$\,GeV.} 
  \label{fig2}
 \end{figure}

In order to track automatically the correct position, the drive system
was modified to use a table providing the angle and amplitude of the Moon shadow
deviation as a function of the azimuth and zenith angle of the
observation. For the 2010--2011 observations, the table was built using
a simple dipole model for the geomagnetic field. The real position of
the Moon shadow can be then slightly off--centred ($<0.3^\circ$) but
this can be corrected afterward during the data analysis using a more
precise model.

In order to preserve the quality of the PMTs, lower high voltages than
standard values are applied. The PMT gain is reduced by a factor of
$\sim$1.5. Then, the telescopes can be operated with a NSB light
$\sim$40 times higher than moonless-night extragalactic NSB. With this
relatevly low gain reduction, the standard PMT calibration method can
be used. The NSB induced by squattered moonlight increases
dramatically with the Moon phase and with the proximity to the
Moon. At 4$^\circ$ from the Moon, the MAGIC telescopes can operate
safely below 50\% Moon phase. 

The fast MAGIC data-acquisition system (2\,GSample/s) associated with
small pixel sizes (0.1$^\circ$) allows a highly performing image
cleaning method \cite{Lombardi11}. Tight time constrains (of the order
of 1\,ns) between neighbour pixels are required. The cleaning levels
were increased until fake signals appear in less than 10\% of
events. The obtained cleaning levels are about twice as high as
the standard ones used for dark night observations. 

The energy threshold achieved after the image cleaning is about
200\,GeV at 30$^\circ$ from zenith. Figure~3 shows the energy
threshold for electrons as a function of the zenith angle. Beyond
50$^\circ$ from zenith, the energy threshold is well above 300\,GeV
and increases dramatically. Thus, we restrained the observation to
zenith angle below 50$^\circ$. As a small phase corresponds to a small
angle between the Moon and Sun, a small-phase Moon is rarely close to
zenith during the night. There are only about 40\,h/year with a
$<$50\%-phase Moon at less than 50$^\circ$ from zenith during the
night. When this happens, the Moon is towards the East (rising just
before the Sun) or towards the West (falling just after the Sun
set). As it is shown Figure~2, in this configuration the Moon shadow
spreads almost vertically above and below the Moon. The shadow above
the moon is in a better position to be observed because closer to
zenith (lower energy threshold) and further aways from the bright side
of the Moon (less background light). Then, we observe the $e^-$ shadow
at the beginning of the nights from December to May and the $e^+$
shadow at the end of the nights from August to January. 

 \begin{figure}[!t]
  \centering
  \includegraphics[width=6.5cm]{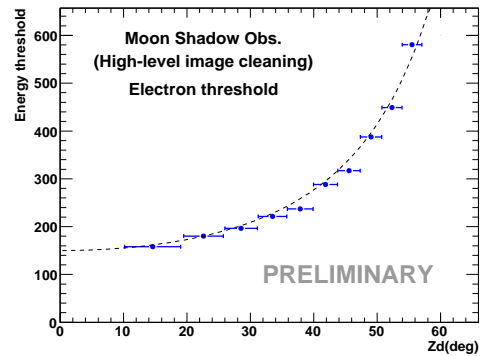}
  \caption{MAGIC energy threshold for electrons during the Moon shadow
    observation as a function of the zenith angle} 
  \label{fig3}
 \end{figure}

\section{Performance and prospect}
In order to check the performance of the MAGIC telescopes, we observed
the Crab Nebula (standard candle of $\gamma$-ray astronomy) in the
same conditions as the Moon shadow. We carried out 2.1\,h of
observation at zenith angle from 8$^\circ$ to 40$^\circ$. Data are
analyzed with the same method as the Moon shadow data (same image
cleaning). Figure~4 shows the $\theta^2$ plot\footnote{$\theta$ is the
  distance between the reconstructed event direction and the source
  position} 
obtained with these data above 300\,GeV. It corresponds to a
sensitivity in 50\,h of about 1.2\% of the Crab Nebula flux. The
angular resolution is almost not degraded by the high NSB level
($<$0.1$^\circ$). It is anyhow much smaller than the Moon radius  
($\sim$0.25$^\circ$).

The shape of the Moon shadow as seen by MAGIC does not depend only on
the angular resolution but also on the energy resolution because the
assumed deviation angle depends on the estimated energy. From
simulation, we estimated the energy resolution to be below 20\% for
diffuse electrons. We expect a deficit extension of about the Moon
size in the direction perpendicular to the deviation axis
(North-South) and 2--3 times larger along the deviation axis
(East-West). For such estended sources, the sensitivity of MAGIC is
reduced by a factor 2--3. Moreover, only the electrons with energy
inside the target range (300-700 GeV) must be counted. Eventually, a
preliminary estimation of the sensitivity of MAGIC for the Moon shadow
is about 4.4\% Crab Unit. 

 \begin{figure}[!t]
  \centering
  \includegraphics[width=6.5cm]{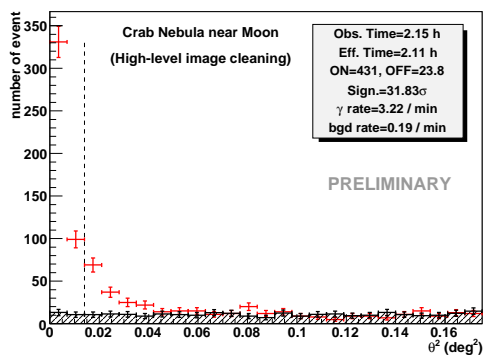}
  \caption{$\theta^2$ distribution ($>$300\,GeV) for the Crab Nebula
    observed with the same NSB light as Moon shadow observation.} 
  \label{fig4}
 \end{figure}

The total missing flux of the electron-positron Moon shadow is the
product of the all-electron spectrum (figure 1) and the solid angle of
the Moon $(6.6\pm0.8)\times 10^{-5}$\,sr which varies of $\pm12$\% as
a function of the observer-Moon distance. This missing flux is shared
between the e$^-$ and the e$^+$ shadows. Table~1 gives the mean
missing flux between 300 and 700\,GeV in Crab unit for the e$^-$ and
the e$^+$ shadows according to several hypothesis. It shows also the
estimated MAGIC observation time required for a detection. In
realistic scenarios, MAGIC would need about 50\,h to detect the e$^-$
shadow and at least 100\,h for the e$^+$ shadow. This is longer than
the available time per year($\sim$20\,h for each shadow). Because of
bad weather, we collected in fact $<$10\,h per shadow during 2010-2011
campaign. With the current observation strategy, MAGIC would need then
several years for a significant detection. 

 \begin{table}[!h]
  \centering
\small{%
  \begin{tabular}{|c|c|c|}
  \hline
   Composition  & Missing flux & Detection time \\
   hypothesis   & 300-700\,GeV & with MAGIC \\
   \hline
   MAGIC spectrum \cite{Borla11}: & &\\
   100\% e-  & 5.4\% & $\sim$30\,h \\
   80\% e-  & 4.3\% & $\sim$50\,h \\
   60\% e-  & 3.3\% & $\sim$90\,h \\
   40\% e+  & 2.2\% & $\sim$200\,h \\
   20\% e+  & 1.1\% & $\sim$800\,h \\
   \hline
   ATIC spectrum \cite{Chang08} : & &\\
   100\% e- & 7.2\% & $\sim$20\,h \\
   80\% e-  & 5.7\% & $\sim$30\,h \\
   60\% e-  & 4.3\% & $\sim$50\,h \\
   40\% e+  & 2.9\% & $\sim$100\,h \\
   20\% e+  & 1.5\% & $\sim$400\,h \\
  \hline
  \end{tabular}
}

  \caption{Mean missing flux of the Moon shadow in Crab nebula
    $\gamma$-ray-flux unit for different composition hypothesis.} 
  \label{table_simple}
  \end{table}

The next generation of IACT (CTA\cite{CTA}) should have a sensitivity
an order of magnitude better than MAGIC. If observation with strong
Moon light is possible, the $e^{-}$ shadow detection time would
decrease dramatically ($<$5\,h). The accessible energy range should
also widen at lower energies thanks to larger telescopes with larger
field of views, and at higher energies thanks to larger effective
collection area.

\section{Conclusion}
Using the Moon shadow effect, IACT arrays have a chance to measure
cosmic ray $e^+$/$e^-$ ratio at energies hardly accessible with
satellite experiments. Probing this ratio at the TeV regime is
particularly interesting because of features were reported in the
all-electron spectrum and because the positron fraction shows an
unpredicted rise above 10\,GeV. However observation near the Moon
is very challenging for IACT because of the high NSB induced by the
scattered moonlight. Using low gain PMT with reduced HV, the MAGIC
collaboration started observation of the electron Moon shadow in the
energy range 300-700\,GeV. These observations carried out at 4$^{\circ}$
from the Moon provide an energy threshold of 200\,GeV and a
sensitivity above 300\,GeV corresponding to 1.2\% Crab Nebula units. In
spite of this good performance, the detection of the electron Moon
shadow with MAGIC will require several years because of the short
observation window available every year.


\begin{thebibliography}{}

 \bibitem{Strong04}  A. W. Strong et al., ApJ 613 (2004), 962
 \bibitem{Abdo09} A. A. Abdo et al., PRL 102 (2009), 181101
 \bibitem{Chang08} J. Chang et al., Nature 456 (2008), 362
 \bibitem{Aharonian08} F. A. Aharonian et al., PRL 101 (2008), 261104
 \bibitem{Borla11} D. Borla Tridon et al., 32$^{nd}$ ICRC, Beijing (2011)

 \bibitem{Adriani09} O. Adriani et al., Nature 458 (2009), 607
 \bibitem{Mocchiutti11} E. Mocchiutti et al., 32$^{nd}$ ICRC, Beijing (2011)
 \bibitem{Beatty04} J. J. Beatty et al., PRL 93 (2004), 241102
 \bibitem{Aguilar07} M. Aguilar et al., Physics Letters B 646 (2007), 145
 \bibitem{Vandenbroucke11} J. Vandenbroucke et al., 32$^{nd}$ ICRC, Beijing (2011)
 \bibitem{Fan10} Y.-Z. Fan, et al., Int. J. Mod. Phys. D19 (2010), 2011

 \bibitem{AMS02} A. Kounine et al., 32$^{nd}$ ICRC, Beijing (2011)
 \bibitem{Colin09} P.~Colin et al., 31$^{st}$ ICRC, {\L}\'{o}d\'{z} (2009), arXiv:0907.1026
 \bibitem{DiSciascio11} G. Di Sciascio et al., 32$^{nd}$ ICRC, Beijing (2011)
 \bibitem{Boersma11} D. Boersma et al., 32$^{nd}$ ICRC, Beijing (2011)
 \bibitem{Riviere11} C. Rivi\`ere et al., 32$^{nd}$ ICRC, Beijing (2011)
 \bibitem{l3c2004} P. Achard et al., Astropart. Phys. 23 (2005), 411
 \bibitem{tibet2007} M. Amenomori et al., Astropart. Phys. 28 (2007), 137
 \bibitem{artemis2001} D. Pomared et al., Astropart. Phys. 14 (2001), 287

 \bibitem{Carmona11} E. Carmona et al., Proc. of 32$^{nd}$ ICRC, Beijing, 2011
 \bibitem{Britzger09} D.~Britzger et al., 31$^{st}$ ICRC, {\L}\'{o}d\'{z} (2009), arXiv:0907.0973.
 \bibitem{Lombardi11} S. Lombardi et al., Proc. of 32$^{nd}$ ICRC, Beijing, 2011
 \bibitem{CTA} M. Martinez et al., Proc. of 32$^{nd}$ ICRC, Beijing, 2011
\end{thebibliography}
\end{document}